\documentclass[11pt]{article}
\usepackage{amsmath}
\usepackage{amsfonts}
\usepackage{amssymb}
\usepackage{graphicx}
\usepackage{bm}
\usepackage{color}
\usepackage{amscd}

\def\bea{\begin{eqnarray}}
\def\eea{\end{eqnarray}}

\begin{document}
\begin{center}
\LARGE {\bf Non-linear regime of the Generalized Minimal Massive Gravity in critical points}
\end{center}
\begin{center}
{\bf M. R. Setare\footnote{rezakord@ipm.ir} }\hspace{1mm} ,
H. Adami \footnote{E-mail: hamed.adami@yahoo.com}\hspace{1.5mm} \\
 {Department of Science, University of  Kurdistan  \\
Sanandaj, IRAN.}
 \\
 \end{center}
\vskip 3cm

\begin{abstract}
The Generalized Minimal Massive Gravity (GMMG) theory is realized
by adding the CS deformation term, the higher derivative deformation term, and an extra term
to pure Einstein gravity with a negative cosmological constant. In the present paper we obtain exact solutions to the GMMG field equations in the non-linear regime of the model. GMMG model about $AdS_3$ space is conjectured to be dual to a 2-dimensional CFT. We study the theory in critical points corresponding to the central charges $c_-=0$ or $c_+=0$, in the non-linear regime. We show that $AdS_3$ wave solutions are present, and have logarithmic form in critical points. Then we study the $AdS_3$ non-linear deformation  solution. Furthermore we obtain logarithmic deformation of extremal BTZ black hole. After that using Abbott-Deser-Tekin method we calculate the energy and angular momentum of these types of black hole solutions.
\end{abstract}

\newpage

\section{Introduction}
Recently an interesting three-dimensional massive gravity was introduced by Bergshoeff et al \cite{17}  and dubbed Minimal Massive Gravity (MMG). Linearization of MMG model about $AdS_3$ space gives identical result to Topological Massive Gravity(TMG) \cite{4}. Furthermore Hamiltonian analysis
of non-linear theory gives some phase-space dimension as TMG. So both models have the same spectrum \cite{18}. However, in contrast to TMG, there is not a bulk vs boundary clash in the framework of this new model. More recently the author of \cite{f} has introduced  Generalized Minimal Massive Gravity (GMMG), an interesting modification of MMG. GMMG is a unification of MMG with New Massive Gravity (NMG) \cite{6}, so this model is realized by adding higher-derivative deformation term to the Lagrangian of MMG. $3D$ gravity models are rather interesting from the point of view of the AdS/CFT correspondence, in particular, the MMG model has attracted attention as a theory that circumvents the difficulty of defining a bulk theory with positive-energy propagating modes which at the same time has a CFT dual with positive central charges. This has been called the ``bulk-boundary unitarity clash'' by the authors of \cite{17}. As has been shown in \cite{f}, GMMG also avoids the aforementioned ``bulk-boundary unitarity clash''.
Calculation of the GMMG action to quadratic order about $AdS_3$ space show that the theory is free of negative-energy bulk modes. Also Hamiltonian analysis show that the GMMG model has no Boulware-Deser ghosts and this model propagate only two physical modes.
 So these models are viable candidate for semi-classical limit of a unitary quantum $3D$ massive gravity.\\
 During last year some interesting exact solutions of MMG have been obtained, such as warped $(A)dS$ \cite{5'}, a two-parameter deformation of extremal BTZ black hole \cite{gir}, a non-BTZ AdS black hole \cite{7'}, Petrov-types $O$, $N$ and $D$ solutions \cite{8'}, and finally Kundt solutions \cite{9'}. Investigation such solutions in the framework of GMMG model would also be
interesting.
It has been shown \cite{f} that GMMG exhibits
not only massless graviton solutions, but also $log$ and $log^2$ solutions. The $log$ and $log^2$ modes appear in tricritical points in the parameter
space of GMMG model. So in such special point in parameter space all massive gravitons become massless. In another terms the massive graviton modes
that satisfy Brown-Henneaux boundary conditions, in the tricritical points replaced by $log$ and $log^2$ solutions, which obey $log$ and $log^2$
boundary conditions toward $AdS_3$ boundary.\\
In the present paper we would like to obtain exact solutions to the GMMG field equations in the non-linear regime of the model. We look for $AdS_3$-wave solutions in $AdS_3$ space. The presence of $AdS_3$-wave solutions which behave as logarithmic modes toward the AdS boundary indicate on the existence of critical points in the parameter space of the GMMG model. In the another term, we show that in the parameter space of GMMG model, not only in the linear level there is logarithmic modes in the critical points \cite{f}, but also in the non-linear level this model has the exact solutions with logarithmic behaviour. We study the theory in critical points corresponding to the central charges $c_-=0$ or $c_+=0$, in the non-linear regime.
\section{Generalized Minimal Massive Gravity}
We introduce the Lagrangian
of GMMG model as \cite{f}
  \begin{equation} \label{2}
 L_{GMMG}=L_{GMG}+\frac{\alpha}{2}e.h\times h
\end{equation}
where
\begin{equation} \label{2'}
 L_{GMG}=L_{TMG}-\frac{1}{m^2}(f.R+\frac{1}{2}e.f\times f)
\end{equation}
here $m$ is mass parameter of NMG term and $f$ is an auxiliary one-form field. $L_{TMG}$ is the Lagrangian of TMG,
 \begin{equation} \label{1'}
 L_{TMG}=-\sigma e.R+\frac{\Lambda_0}{6} e.e\times e+h.T(\omega)+\frac{1}{2\mu}(\omega.d\omega+\frac{1}{3}\omega.\omega\times\omega)
\end{equation}
 where $\Lambda_0$ is a cosmological parameter with dimension of mass squared, $\sigma$ is a sign, $\mu$ is the mass parameter of the Lorentz Chern-Simons term, $\alpha$ is a dimensionless parameter, $e$ is a dreibein, $h$ is the auxiliary field, $\omega$ is a dualized  spin-connection, and $T(\omega)$ and
  $R(\omega)$ are the Lorentz covariant torsion and the curvature 2-form, respectively. So by adding extra term $\frac{\alpha}{2}e.h\times h$ to the Lagrangian of generalized massive gravity we obtain Lagrangian of GMMG model. The equation for metric can be obtained by generalizing  field equation of MMG. Now we introduce GMMG field equation
as follows \cite{f}
\begin{equation}\label{81}
\Phi ^{\mu \nu} = \Lambda _{0} g^{\mu \nu} + \sigma G^{\mu \nu} + \frac{1}{\mu} C^{\mu \nu} + \frac{\gamma}{\mu ^{2}}
J^{\mu \nu} + \frac{s}{2 m^{2}} K^{\mu \nu}=0 ,
\end{equation}
where $ G ^{\mu \nu}$ is Einstein's tensor, the Cotton tensor is
\begin{equation}\label{91}
C^{\mu \nu}=\frac{1}{\sqrt{-g}} \varepsilon ^{\mu \alpha \beta} \nabla _{\alpha} S^{\nu}_{\beta},
\end{equation}
where $ S^{\mu}_{\nu}=R^{\mu}_{\nu}-\frac{1}{4}\delta^{\mu}_{\nu}R $ is the Schouten tensor in 3 dimensions,
\begin{equation}\label{101}
J^{\mu \nu} =\frac{1}{2g}  \varepsilon ^{\mu \rho \sigma}  \varepsilon ^{\nu \alpha \beta} S_{\rho \alpha} S_{\sigma \beta},
\end{equation}
and
\begin{equation}\label{111}
K^{\mu \nu}=2 \Box R^{\mu \nu} - \frac{1}{2} \nabla ^{\mu} \nabla ^{\nu} R - \frac{1}{2} g^{\mu \nu} \Box R - 8 R^{\mu \alpha}
R^{\nu}_{\alpha}+\frac{9}{2} R R^{\mu \nu} + 3 g^{\mu \nu}  R^{\alpha \beta} R_{\alpha \beta} - \frac{13}{8} g^{\mu \nu} R^{2} ,
\end{equation}
$s$ is sign, $\gamma$, $\sigma$ $\Lambda_{0}$ are the parameters
 which defined in terms of cosmological constant $\Lambda=\frac{-1}{l^2}$, $m$, $\mu$, and the sign of Einstein-Hilbert term.
 The field equation (\ref{81}) admits $AdS_{3}$ solution,
\[
d\bar{s}^2=l^{2}(-\cosh ^{2}\rho ~d\tau ^{2}+\sinh
^{2}\rho ~d\phi ^{2}+d\rho ^{2})
\]
where $l^{2}\equiv -\Lambda ^{-1}$ fixes with parameters of theory.
The field equation for $AdS_{3}$ reduces to an quadratic equation for $%
\Lambda $
\begin{equation}
\left( \frac{\gamma }{4\mu ^{2}}-\frac{s}{4m^{2}}\right) \Lambda
^{2}-\Lambda \sigma+\Lambda_{0}=0.  \label{AdS Constraint}
\end{equation}
So,
\begin{equation}  \label{51}
\Lambda=\frac{(\sigma\pm\sqrt{\sigma^{2}-\Lambda_{0}(\frac{\gamma}{\mu^2}-\frac{s}{m^2})})}{\frac{1}{2}
(\frac{\gamma}{\mu^2}-\frac{s}{m^2})}
\end{equation}
GMMG model about $AdS_3$ space is conjectured to be dual to a 2-dimensional CFT  with following central charges,\footnote{In our notation $G=1$.} \cite{f}
\begin{equation}\label{5}
c_{-}=\frac{3l}{2}(\sigma +\frac{\gamma}{2 \mu ^{2} l^{2}}+\frac{s}{2 m^{2} l^{2}}-\frac{1}{\mu l}) ,\hspace{1cm} c_{+}=\frac{3l}{2}(\sigma +\frac{\gamma}{2 \mu ^{2} l^{2}}+\frac{s}{2 m^{2} l^{2}}+\frac{1}{\mu l}),
\end{equation}
As has been discussed in \cite{f}, we have a couple critical points in the space of parameters, where in first critical point $c_{-}=0$, so
\begin{equation}\label{201}
 \sigma=\frac{1}{\mu l}-\frac{s}{2m^{2}l^{2}}-\frac{\gamma }{2 \mu ^{2}l^2}
\end{equation}
In the second critical point $c_{+}=0$, then
\begin{equation}\label{261}
 \sigma=-\left( \frac{\gamma }{2 \mu ^{2}l^2} + \frac{s }{2m^{2}l^2}+\frac{1}{\mu l} \right)
\end{equation}
Here we are interested to critical points $c_{-}=0$ and $c_{+}=0$, separately.

\section{AdS$_{3}$-waves solutions}
In this section we are looking for AdS$_{3}$-wave solutions in AdS$_{3}$ space, also known as pp-waves.
AdS$_{3}$-wave solutions in the framework of TMG and MMG  have been studied in  \cite{6'}, \cite{gir} respectively.
Actually the AdS$_{3}$-waves are a special class of spacetimes defined when the negative cosmological constant is present.
This class is called Siklos spacetimes \cite{siklos} and their special feature is the presence of a multiple
principal null-directed Killing vector  $k^\mu$ of their Weyl tensor.
 For our purpose, let us consider AdS$_{3}$ metric written in Poincare coordinates,
\begin{equation}\label{6}
d \bar{s} ^{2}=\bar{g} _{\mu \nu} dx^{\mu} dx^{\nu}=\frac{l^{2}}{z^{2}} \left( -2dx^{+}dx^{-}+dz^{2} \right) ,
\end{equation}
 where$ z \in [0,\infty )$ while $x^{\pm} \in \mathbb{R}$ and the boundary of the space is located at $z=0$. Also, in the above metric, the AdS$_{3}$ radius $l$ is a positive constant. Now, we consider the Kerr-Schild deformation of AdS$_{3}$ spacetime as
 \begin{equation}\label{01'}
   g_{\mu \nu} = \bar{g} _{\mu \nu} - F(x^{+},z) k_{\mu} k_{\nu},
 \end{equation}
where the $F(x^{+},z)$ is the wave profile and $k_{\mu}$ is a null, tangent to geodesic vector field. If we suppose that $k=\frac{z}{l} \partial _{x^{+}}$ then we have the AdS$_{3}$-wave metric as follows:
\begin{equation}\label{7}
ds^{2}=\frac{l^{2}}{z^{2}} \left( -F(x^{+},z)(dx^{+})^{2}-2dx^{+}dx^{-}+dz^{2} \right) .
\end{equation}
Now, we demand that this metric be a solution to GMMG field equation. To this end, we must have,
\begin{equation}\label{8}
\Lambda _{0} + \frac{\sigma}{l^2} + \frac{\gamma }{4 \mu ^{2} l ^{4}} - \frac{s }{4 m^{2}  l ^{4}}=0 ,
\end{equation}
which is exactly relation (\ref {AdS Constraint}), and this is natural because the AdS$_{3}$-wave solution (\ref{7}) exhibit the gravitational waves propagating on the AdS$_{3}$ background.
On the other hand, $F(x^{+},z)$ must satisfy the following partial differential equation (PDE),
$$ (2 \sigma \mu ^{2} l^{2} m^{2} + \gamma m^{2} - s \mu ^{2}) \dfrac{\partial F(x^{+},z)}{\partial z} - z (2 \sigma \mu ^{2} l^{2} m^{2} + \gamma m^{2}  - s \mu ^{2}) \dfrac{\partial ^{2} F(x^{+},z)}{\partial z ^{2}}$$
\begin{equation}\label{9}
 + 2 \mu z^{2} (-2s\mu +l m^{2}) \dfrac{\partial ^{3} F(x^{+},z)}{\partial z ^{3}} - 2 s \mu ^{2} z^{3} \dfrac{\partial ^{4} F(x^{+},z)}{\partial z ^{4}}=0.
\end{equation}
If we divide the above PDE by $m^{2}$ then by taking $m^{2} \rightarrow \infty $ we will obtain following PDE which previously has obtained  for MMG model \cite{gir}
\begin{equation}\label{02'}
      (2 \sigma \mu ^{2} l^{2} + \gamma ) \dfrac{\partial F(x^{+},z)}{\partial z} - z (2 \sigma \mu ^{2} l^{2} + \gamma ) \dfrac{\partial ^{2} F(x^{+},z)}{\partial z ^{2}} + 2 l \mu z^{2} \dfrac{\partial ^{3} F(x^{+},z)}{\partial z ^{3}} =0.
\end{equation}
Therefore, by comparing \eqref{9} and \eqref{02'}, we see that the fourth order derivative in \eqref{9} is new term and it is due to the contribution of NMG part that appear  in GMMG model.
The generic solution of this differential equation is as follows:
\begin{equation}\label{10}
F(x^{+},z)=F_{0}(x^{+})+F_{2}(x^{+})z^{2}+F_{\alpha _{+}}(x^{+})z^{\alpha _{+}}+F_{\alpha _{-}}(x^{+})z^{\alpha _{-}},
\end{equation}
where
\begin{equation}\label{11}
\alpha _{\pm}= 1+\frac{l m^{2}}{2 s \mu} \pm \sqrt{\left( \frac{l m^{2}}{2 s \mu} \right)^{2} - \frac{1}{2 s \mu ^{2}}\left( 2 \sigma \mu ^{2} l^{2} m^{2} + \gamma m^{2} - s \mu ^{2} \right)}  .
\end{equation}
Two special cases can be considered: (i) If parameters of this model are selected such that $ 2 \sigma \mu ^{2} l^{2} m^{2} + \gamma m^{2} - s \mu ^{2} =0 $ then the equation \eqref{11} becomes
$\alpha _{+}= 1+\frac{l m^{2}}{ s \mu}$ and $\alpha _{-}= 1$ . (ii) If we take parameters of this model such that $ 1+\frac{l m^{2}}{2 s \mu} = 0$ then \eqref{11} reduces to
\begin{equation}\label{03'}
\alpha _{\pm}= \pm \sqrt{ 2 \sigma \mu l +\frac{\gamma}{\mu l} + \frac{3}{2}}  .
\end{equation}
As we have already mentioned in the introduction and have been shown in \cite{f}, there are special points
in the parametric space which correspond to solutions with logarithmic
terms and which are associated to a relaxed set of boundary
conditions. In the framework of MMG,
$AdS_3$ wave solutions with a logarithmic wave profile have been obtained in \cite{gir} (see also \cite{Giribet:2010ed,Grumiller:2009mw,set3}).
So in addition to the above result, we can have three other solutions. One of them can be found at critical point $c_{-}=0$. If we solve PDE (\ref{9}) at this critical point, we get the following result,
\begin{equation}\label{12}
F(x^{+},z)=F ^{(c_{-})} _{0}(x^{+})+F ^{(c_{-})} _{1}(x^{+})z^{2}+F ^{(c_{-})} _{2}(x^{+})z^{2} \ln (z) +F ^{(c_{-})} _{3}(x^{+})z^{\frac{l m^{2}}{s \mu}},
\end{equation}
Another solution come from solving the equation (\ref{9}) at critical point $c_{+}=0$ and it can be expressed as follow:
\begin{equation}\label{13}
F(x^{+},z)=F ^{(c_{+})} _{0}(x^{+})+F ^{(c_{+})} _{1}(x^{+})z^{2}+F ^{(c_{+})} _{2}(x^{+}) \ln (z) +F ^{(c_{+})} _{3}(x^{+})z^{2+\frac{l m^{2}}{s \mu}},
\end{equation}
As the last special solution, we consider a case in which, in addition to $c_{+}=0$, we have $\frac{l m^{2}}{2s \mu}=-1 $, then,in this case the solution is given by
\begin{equation}\label{14}
F(x^{+},z)= \tilde{F} _{0}(x^{+})+ \tilde{F} _{1}(x^{+})z^{2}+ \tilde{F} _{2}(x^{+}) \ln{z}+ \tilde{F} _{3}(x^{+})(\ln{z})^{2}.
\end{equation}
So we have obtained AdS waves that exhibit logarithmic asymptotic behavior at critical points. The existence of these logarithmic modes can be seen as a first sign that critical GMMG models are dual to logarithmic conformal field theories, when appropriate boundary conditions are imposed (see also \cite{go}).

\section{AdS$_{3}$ non-linear deformed solutions}
Let us begin this section by applying following coordinate transformation on (\ref{6}),
\begin{equation}\label{15}
r=\sqrt{2} l^{2}/z, \hspace{2 cm} x^{\pm}=t \pm l \phi ,
\end{equation}
so the AdS$_{3}$ metric can be rewritten as follows:
\begin{equation}\label{16}
ds_{0}^{2}=-\frac{r^{2}}{l^{2}}dt^{2}+\frac{l^{2}}{r^{2}}dr^{2}+r^{2}d \phi ^{2},
\end{equation}
where the equation (\ref{8}) still is valid. In similar way with the previous section, we introduce a deformation of the form
\begin{equation}\label{17}
ds^{2}=ds_{0}^{2}+H_{-}(t,r)(dt - l d\phi)^{2}+H_{+}(t,r)(dt + l d\phi)^{2},
\end{equation}
Now, we are looking for time-independent solutions. For this purpose, first we suppose that
\begin{equation}\label{18}
H_{+}(t,r)=0 \hspace{1cm} and \hspace{1cm} H_{-}(t,r)=H_{-}(r),
\end{equation}
and we try to find a solution at critical point $c_{-}=0$. For this case then we find the following solutions
\begin{equation}\label{19}
H_{-}(r)=C_{0}+C_{1} r^{2}+C_{2} \ln{r} +C_{3} r^{2-\frac{l m^{2}}{s \mu}} ,
\end{equation}
where $C_{i}$'s are constants of integration. This solution behaves as the Log Gravity excitation asymptotically \cite{gir,7,9}, and exhibit non-linear realization of GMMG model.
 We can find another solution when we take (\ref{18}) and the critical point $c_{+}=0$, so the solution for these assumptions is:
\begin{equation}\label{20}
H_{-}(r)=C_{0}+C_{1} r^{2}+C_{2} r^{2} \ln{r} +C_{3} r^{-\frac{l m^{2}}{s \mu}} ,
\end{equation}
For assumption (\ref{18}), we can find $\log ^{2}$ solutions by taking $c_{-}=0$ and $\frac{l m^{2}}{2s \mu}=1$. So we obtain
\begin{equation}\label{21}
H_{-}(r)=C_{0}+C_{1} r^{2}+C_{2} \ln{r} +C_{3} (\ln{r})^{2}.
\end{equation}
On the other hands, for the case
\begin{equation}\label{22}
H_{-}(t,r)=0 \hspace{1cm} and \hspace{1cm} H_{+}(t,r)=H_{+}(r),
\end{equation}
, one can find that
\begin{equation}\label{23}
H_{+}(r)=C_{0}+C_{1} r^{2}+C_{2} \ln{r} +C_{3} r^{2+\frac{l m^{2}}{s \mu}} \hspace{1.5cm} at \hspace{0.5cm} c_{+}=0,
\end{equation}
\begin{equation}\label{24}
H_{+}(r)=C_{0}+C_{1} r^{2}+C_{2} r^{2} \ln{r} +C_{3} r^{\frac{l m^{2}}{s \mu}} \hspace{1.5cm} at \hspace{0.5cm} c_{-}=0,
\end{equation}
and
\begin{equation}\label{25}
H_{+}(r)=C_{0}+C_{1} r^{2}+C_{2} \ln{r} +C_{3} (\ln{r})^{2}  \hspace{1cm} at \hspace{0.5cm} c_{+}=0 \hspace{0.5cm} with \hspace{0.5cm} \frac{l m^{2}}{s \mu}=-1.
\end{equation}
 More than $\log^1$, here we have obtained the solutions with $\log^2$ fall-off behavior. For boundary conditions that keep all $log^n$ modes, the two-point functions of the dual CFT are correspond to those
of a rank $n+1$ LCFT \cite{mer}. So the existence and properties of
 $\log^2$ solutions lead one to conjecture that the CFT-dual of this gravity model is
an LCFT of rank 3, if appropriate boundary conditions that include excitations with
$log^2$ boundary behavior are adopted (see also \cite{zo}).

\section{Deformations of eBTZ solution}
In this section we want to generalize previous solutions, and study the deformation of the extremal BTZ (eBTZ) black hole.
Consider the following metric
\begin{equation}\label{26}
ds^{2}=ds^{2}_{eBTZ}++H_{-}(t,r)(dt - l d\phi)^{2}+H_{+}(t,r)(dt + l d\phi)^{2},
\end{equation}
where $ds^{2}_{eBTZ}$ is the extremal BTZ metric which is expressed as follows:
\begin{equation}\label{27}
ds^{2}_{eBTZ}=-\frac{(r^{2}-r_{e}^{2})^{2}}{l^{2} r^{2}} dt^{2} +
 \frac{l^{2} r^{2}}{(r^{2}-r_{e}^{2})^{2}}dr^{2} + r^{2} \left( d\phi - \frac{r_{e}^{2}}{l r^{2}}dt \right)^{2} ,
\end{equation}
where $r_{e}$ is the extremal black hole radii. As previous section, we can classify time-independent deformed eBTZ solutions as follow. First, we consider the case of (\ref{18}), thus we have
\begin{equation}\label{28}
H_{-}(r)=a_{0}+a_{1} (r^{2}-r_{e}^{2})+a_{2} \ln{(r^{2}-r_{e}^{2})} +a_{3} (r^{2}-r_{e}^{2})^{1-\frac{l m^{2}}{2s \mu}} \hspace{1.5cm} at \hspace{0.5cm} c_{-}=0,
\end{equation}
\begin{equation}\label{29}
H_{-}(r)=b_{0}+b_{1} (r^{2}-r_{e}^{2})+b_{2} (r^{2}-r_{e}^{2}) \ln{(r^{2}-r_{e}^{2})} +b_{3} (r^{2}-r_{e}^{2})^{-\frac{l m^{2}}{2s \mu}} \hspace{1.5cm} at \hspace{0.5cm} c_{+}=0,
\end{equation}
and
\begin{equation}\label{30}
H_{-}(r)=c_{0}+c_{1} (r^{2}-r_{e}^{2})+c_{2} \ln{(r^{2}-r_{e}^{2})} +c_{3} (\ln{(r^{2}-r_{e}^{2})})^{2} \hspace{0.3 cm} at \hspace{0.3 cm} c_{-}=0 \hspace{0.3 cm} with \hspace{0.3 cm} \frac{l m^{2}}{2s \mu}=1 .
\end{equation}
As a time-dependent solution, we offer an ansatz with the form
\begin{equation}\label{31}
\begin{split}
H_{-}(t,r)=d_{0}+d_{1} (r^{2}-r_{e}^{2})+d_{2} \ln{(r^{2}-r_{e}^{2})}+d_{3} (r^{2}-r_{e}^{2})^{1-\frac{l m^{2}}{2s \mu}} \\
+ \left( k+ \frac{1}{A (r^{2}-r_{e}^{2})^{\frac{l m^{2}}{2s \mu}-1}} \right) t -\frac{k^{2} l^{6}}{B(r^{2}-r_{e}^{2})^{2}} ,\hspace{1.5cm}
\end{split}
\end{equation}
at critical point $c_{-}=0$ and providing $H_{+}(t,r)=0$, this solves the field equations (\ref{81}) when the coefficients $A$ and $B$ are given as
\begin{equation}\label{32}
A=\frac{l^{2} m^{4} (\gamma m^{2}-3 \mu ^{2} s)}{\mu s (l^{2}m^{4} \mu-4l \gamma m^{4} +6ls \mu ^{2} m^{2} + 8 s^2 \mu ^{3})},\hspace{1cm} B=\frac{96 \mu (l m^{2}-6s\mu)}{(2\gamma m^{2}+\mu l m^{2}-8 s \mu^{2})} ,
\end{equation}
and $k$ is an arbitrary constant, like $d_{i}$'s. At critical point $c_{-}=0$ with an additional requirements $\frac{l m^{2}}{2s \mu}=1$, we can propose another ansatz as time-dependent solution which is given by
\begin{equation}\label{33}
\begin{split}
H_{-}(t,r)=\alpha _{0} +\left[ \alpha _{1}+\alpha _{2} (r^{2}-r_{e}^{2}) \right] t -\frac{\alpha _{1} \alpha _{2} l^{6}}{16 (r^{2}-r_{e}^{2})} +\frac{\alpha _{1}^{2} l^{5} (2\gamma - 3 \mu l)}{192 \mu (r^{2}-r_{e}^{2})^{2}} \\
+\alpha _{3} \ln{(r^{2}-r_{e}^{2})} +\alpha _{4} (\ln{(r^{2}-r_{e}^{2})})^{2},\hspace{2.5cm}\\
H_{+}(t,r)=0, \hspace{9.4cm}
\end{split}
\end{equation}
The existence of solutions such as \eqref{31} and \eqref{33} show that GMMG in $AdS_3$ space at critical points have solutions not being Einstein manifolds. Similar time-dependent solutions in $AdS_3$ space at critical point of MMG model have obtained in \cite{gir}. To see about the existence and importance of this type of solutions in the TMG refer to \cite{20,21}, where the authors have discussed that this type of solutions may be relevant for TMG model when one calculate the partition function of Chiral Gravity.
\section{The Conserved Charges of GMMG}
In this section we want to obtain the conserved charges associated to the solutions we obtained above. There are several approach to obtain mass and angular momentum in $3D$ massive gravity. Here we follow the method given by Abbott, Deser, and Tekin in \cite{1,2,3}, which needs to obtain the field equations and linearize them about the (A)dS vacuum of the model. In paper \cite{set} we have obtained the formula for the calculation of conserved charges in GMMG model in asymptotically AdS$_3$ spacetime. Here at first we review some of the results of \cite{set}.
Given the field equations of the theory as
\begin{equation}\label{34}
\Phi   _{\mu \nu} (g, R, \nabla (Riemann) , R ^{2} , ....) = 8 \pi T _{\mu \nu}.
\end{equation}
We assume that $\Phi _{\mu \nu} $ is covariantly
conserved, so $\nabla _{\mu} \Phi ^{\mu \nu} = 0$. We consider the metric $\bar{g}_{\mu \nu}$ as background metric which solves following equation
\begin{equation}\label{35}
\Phi _{\mu \nu} (\bar{g}_{\mu \nu}) =0,
\end{equation}
 so $\bar{g}_{\mu \nu}$ is vacuum solution of the above field equations. Now, we demand that $g_{\mu \nu}=\bar{g}_{\mu \nu} + h_{\mu \nu}$ also be a solution for (\ref{34}). Thus, for linear part of $\Phi ^{\mu \nu} $ with respect to $ h_{\mu \nu}$, we have
\begin{equation}\label{36}
\nabla _{\mu} \Phi _{L}^{\mu \nu}=0.
\end{equation}
We suppose that $\bar{\xi} ^{\mu}$ be a killing vector field that the spacetime in question admits it such as background spacetime. Then using the killing equation, one can show that $\kappa ^{\mu} = \bar{\xi} _{\nu} \Phi _{L}^{\mu \nu} $ is a conserved current, that is,$\nabla _{\mu} \kappa ^{\mu}=0$ and thus we can rewrite $\kappa ^{\mu}$ as follow \cite{3}:
\begin{equation}\label{37}
\kappa ^{\mu}=\nabla _{\nu} \mathcal{F}^{\mu \nu}.
\end{equation}
Therefore, we can define conserved charge associate to Killing vector field $\bar{\xi} ^{\mu}$ which is given by
\begin{equation}\label{38}
Q(\xi)=c \int _{\mathcal{M}} \sqrt{-g} \bar{\xi} _{\nu} T^{\mu \nu} dS_{\mu} = \frac{c}{16 \pi} \int _{\partial \mathcal{M}} \sqrt{-g} \mathcal{F}^{\mu \nu}(\bar{\xi}) dS_{\mu \nu} ,
\end{equation}
where $\partial \mathcal{M}$ is boundary of the hypersurface $\mathcal{M}$. Since we are interested in solutions that are asymptotically AdS$_{3}$ (which takes as the background spacetime) and they can be written as  $g_{\mu \nu}=\bar{g}_{\mu \nu} + h_{\mu \nu}$ asymptotically, where $\bar{g}_{\mu \nu}$ is the AdS$_{3}$ spacetime metric, so we can take $\mathcal{M}$ as a space-like hypersurface whose boundary located at spatial infinity. In this manner, by working in the coordinates that AdS$_{3}$ matric can be written in the form (\ref{16}), if we take $\bar{\xi}^{\mu}=\partial _{t}$ and $c=-4$, we will obtain ADM energy, on the other hands, if we take  $\bar{\xi}^{\mu}=\partial _{\phi}$ and $c=4$, we will have angular momentum of considered spacetime. Strictly speaking,
\begin{equation}\label{39}
E=-\frac{1}{2 \pi} \int _{\partial \mathcal{M}} \sqrt{-g} \mathcal{F}^{0 i}(\partial _{t}) dS_{i} ,
\end{equation}
and
\begin{equation}\label{40}
J= \frac{1}{2 \pi} \int _{\partial \mathcal{M}} \sqrt{-g} \mathcal{F}^{0 i}(\partial _{\phi}) dS_{i} ,
\end{equation}
where $dS_{i}=dS_{0i}$ is the line element on $\partial \mathcal{M}$. Now, assume that $\Phi ^{\mu \nu} $ in formula (\ref{34}) is given by (\ref{81}). In general, $\Phi ^{\mu \nu} $ does not satisfied Bianchi identity, but fortunately this happens for background AdS$_{3}$ spacetime and we can use the above procedure for calculating conserved charges of considered solutions. For this field equations the authors in \cite{set} demonstrated that the superpotential $ \mathcal{F}^{\mu \nu}(\bar{\xi}) $ for solutions that are asymptotically AdS$_{3}$ is given by
\begin{equation}\label{41}
\mathcal{F}^{\mu \nu}(\bar{\xi})=\left( \sigma +\frac{\gamma }{2 \mu ^{2} l^{2}} + \frac{s }{2m^{2} l^{2}} \right) q_{E}^{\mu \nu} (\bar{\xi})+\frac{1}{2 \mu} q_{E}^{\mu \nu} (\bar{\Xi}) + \frac{1}{2 \mu} q_{C}^{\mu \nu} (\bar{\xi}) +\frac{s}{2m^{2}} q_{N}^{\mu \nu} (\bar{\xi}),
\end{equation}
where
$$q_{E}^{\mu \nu} (\bar{\xi}) = 2 \sqrt{-\bar{g}} \left( \bar{\xi}_{\lambda}  \bar{\nabla}^{ [ \mu} h^{\nu ] \lambda}
 + \bar{\xi}^{ [ \mu}  \bar{\nabla}^{\nu ] } h + h^{\lambda [ \mu}  \bar{\nabla}^{\nu ] } \bar{\xi}_{\lambda} +
\bar{\xi} ^{ [ \nu} \bar{\nabla} _{ \lambda} h^{\mu ] \lambda} + \frac{1}{2} h \bar{\nabla} ^{ \mu} \bar{\xi}^{\nu} \right) ,$$
$$q_{C}^{\mu \nu} (\bar{\xi}) = \varepsilon ^{\mu \nu}  _{\hspace{3 mm} \alpha} \mathcal{G}_{L}^{\alpha \beta} \bar{\xi}_{\beta}
 + \varepsilon ^{\beta \nu} _{\hspace{3 mm} \alpha} \mathcal{G}_{L}^{\mu \alpha} \bar{\xi}_{\beta}
 +\varepsilon ^{\mu \beta}  _{\hspace{3 mm} \alpha} \mathcal{G}_{L}^{\alpha \nu} \bar{\xi}_{\beta} ,$$
 \begin{equation}\label{42}
q_{N}^{\mu \nu} (\bar{\xi}) = \sqrt{-\bar{g}} \left[ 4 \left(  \bar{\xi}_{\lambda}  \bar{\nabla}^{ [ \nu} \mathcal{G}_{L}^{\mu ] \lambda}
+\mathcal{G}_{L}^{\lambda [ \nu} \bar{\nabla} ^{ \mu ] } \bar{\xi}_{\lambda} \right) + \bar{\xi}^{ [ \mu}  \bar{\nabla}^{\nu ] } R_{L}
+ \frac{1}{2} R_{L} \bar{\nabla} ^{ \mu} \bar{\xi}^{\nu} \right] ,
\end{equation}
also, $ \bar{\Xi}^{\beta}=\frac{1}{\sqrt{-\bar{g}}} \varepsilon^{\alpha \lambda \beta} \bar{\nabla} _{ \alpha} \bar{\xi}_{\lambda}  $.
Now, we apply this formalism to calculate conserved charges of the solutions of the type (\ref{26}). At spatial infinity, i.e. for $r\gg r_{e}$, one can read $h_{\mu \nu}$ from (\ref{26}) as follow:
$$h_{tt}=\frac{2 r_{e}^{2}}{l^{2}}+H_{-}(t,r)+H_{+}(t,r),\hspace{0.5 cm}
 h_{t \phi }=-\frac{r_{e}^{2}}{l}-l H_{-}(t,r)+l H_{+}(t,r),$$
\begin{equation}\label{43}
h_{rr}=\frac{2 l^{2} r_{e}^{2}}{r^{4}}, \hspace{0.5 cm} h_{\phi \phi }=l^{2} ( H_{-}(t,r)+H_{+}(t,r) ),
\end{equation}
and the rest are zero. Here, it should be noted that $H_{\pm}$ calculated at spatial infinity. Also, note that we take $\partial \mathcal{M} $ as a circle of infinite radius. By substituting (\ref{43}) into (\ref{41}) and inserting obtained result in (\ref{39})-(\ref{40}) one can find that:
\begin{equation}\label{44}
\begin{split}
 E =\frac{4}{3l}\lim _{r \to \infty} \biggl\{  \left( \frac{r_{e}^{2}}{l^2}+H_{-}-\frac{1}{2}r\dfrac{\partial H_{-}}{\partial r} \right) c_{-} + \left( H_{+}-\frac{1}{2}r\dfrac{\partial H_{+}}{\partial r} \right) c_{+} \\
+\frac{3r}{8 \mu} \left( \dfrac{\partial H_{-}}{\partial r} - \dfrac{\partial H_{+}}{\partial r} -r \dfrac{\partial ^{2} H_{-}}{\partial r^{2}}+r \dfrac{\partial ^{2} H_{+}}{\partial r^{2}}\right) \biggl\} ,
\end{split}
\end{equation}
\begin{equation}\label{45}
\begin{split}
 J=\frac{4}{3}\lim _{r \to \infty}  \biggl\{  \left( \frac{r_{e}^{2}}{l^2}+H_{-}-\frac{1}{2}r\dfrac{\partial H_{-}}{\partial r} \right) c_{-} - \left( H_{+}-\frac{1}{2}r\dfrac{\partial H_{+}}{\partial r} \right) c_{+} \hspace{1cm} \\
 +\frac{3r}{8 \mu} \left( \dfrac{\partial H_{-}}{\partial r} + \dfrac{\partial H_{+}}{\partial r} -r \dfrac{\partial ^{2} H_{-}}{\partial r^{2}}-r \dfrac{\partial ^{2} H_{+}}{\partial r^{2}}\right) +\frac{3l^{4}}{16 \mu r^{2}} \left( \dfrac{\partial ^{2} H_{-}}{\partial t^{2}}+ \dfrac{\partial ^{2} H_{+}}{\partial t^{2}}\right) \biggl\} ,
 \end{split}
\end{equation}
Here, we stated that these formulas are usable just for solutions for which inside of the brackets as $r$ tends to infinity, becomes finite and otherwise this formulas will be useless.\footnote{Here we should mention that may be exist other solutions which are not investigated here. Boundary condition restrict the scope of solutions. So by some special boundary conditions some of solutions will be excluded. To clarifying the matter, we focus on the solution (\ref{31}). If we demand following boundary conditions
$$ g_{\mu \nu}=
\begin{pmatrix}
    \mathcal{O} (r^{2})       & \mathcal{O} (1) & \mathcal{O} (r^{2}) \\
           & \mathcal{O} (\frac{1}{r^{4}}) & \mathcal{O} (1) \\
         &     &  \mathcal{O} (r^{2})
\end{pmatrix}
$$
 $d_{2}$ vanishes in (\ref{31}) and we lost logarithmic term. In any case we must impose some boundary conditions, but they needs to satisfy finiteness of
asymptotic charges. Due to this, here (\ref{28}) and (\ref{31}) are good solutions with finite asymptotic charges. In the other hand there are some interesting solutions, like (\ref{29}), (\ref{30}), etc, whose asymptotic charges can not be calculated by the ADT method.} We now discuss about some special case of the above results. As the first case, we consider a solution for which $H_{+}(t,r)=0$ and $c_{-}=0$, so we have
\begin{equation}\label{46}
 E _{-+}=\frac{1}{2 \mu l}\lim _{r \to \infty} \left( r \dfrac{\partial H_{-}}{\partial r} -r^{2} \dfrac{\partial ^{2} H_{-}}{\partial r^{2}} \right) ,
\end{equation}
\begin{equation}\label{47}
 J_{-+}=\frac{1}{2 \mu}\lim _{r \to \infty}  \left( r\dfrac{\partial H_{-}}{\partial r} -r^{2} \dfrac{\partial ^{2} H_{-}}{\partial r^{2}}+\frac{l^{4}}{2 r^{2}} \dfrac{\partial ^{2} H_{-}}{\partial t^{2}} \right) .
\end{equation}
Using the above formulas, one can calculate energy and angular momentum of the solution (\ref{28}), assuming $\frac{m^{2} l}{2s \mu} > 1$, we have
\begin{equation}\label{48}
 E _{-+}=\frac{2 a_{2}}{\mu l} ,\hspace{2cm} J_{-+}=\frac{2 a_{2}}{\mu}
\end{equation}
As one can see from the above equations for energy and angular momentum, they are proportional to the $a_2$, which is the coefficient of logarithmic term in Eq.(\ref{28}).\\
Energy and angular momentum of the solution (\ref{31}), by requiring $\frac{m^{2} l}{2s \mu} > 1$, can be calculated by Eq.(\ref{46}) and Eq.(\ref{47}) respectively. In this case we obtain
\begin{equation}\label{49}
 E _{-+}=\frac{2 d_{2}}{\mu l} ,\hspace{2cm} J_{-+}=\frac{2 d_{2}}{\mu}
\end{equation}
Similar to the previous case, here we see again that the energy and angular momentum are proportional to the $d_2$, which is the coefficient of logarithmic term in Eq.(\ref{31}). Thus, in these cases, only the contribution of logarithmic terms appear in Energy and angular momentum and the contribution of others terms in solutions (\ref{28}), (\ref{31})  become zero. More than this, it is clear that we can write $ J_{-+} = l E _{-+}$ for both cases.\\
As the second case, we suppose that $H_{+}(t,r)=0$ and $c_{+}=0$, then (\ref{44}),(\ref{45}) reduced to following equations respectively
\begin{equation}\label{50}
 E_{--} =\frac{4}{3l}\lim _{r \to \infty} \biggl\{  \left( \frac{r_{e}^{2}}{l^2}+H_{-}-\frac{1}{2}r\dfrac{\partial H_{-}}{\partial r} \right) c_{-} +\frac{3r}{8 \mu} \left( \dfrac{\partial H_{-}}{\partial r}  -r \dfrac{\partial ^{2} H_{-}}{\partial r^{2}} \right) \biggl\} ,
\end{equation}
\begin{equation}\label{51}
 J_{--}=\frac{4}{3}\lim _{r \to \infty}  \biggl\{  \left( \frac{r_{e}^{2}}{l^2}+H_{-}-\frac{1}{2}r\dfrac{\partial H_{-}}{\partial r} \right) c_{-}
 +\frac{3r}{8 \mu} \left( \dfrac{\partial H_{-}}{\partial r} -r \dfrac{\partial ^{2} H_{-}}{\partial r^{2}} \right) +\frac{3l^{4}}{16 \mu r^{2}} \dfrac{\partial ^{2} H_{-}}{\partial t^{2}} \biggl\} .
\end{equation}
Now, we consider $H_{-}(t,r)=0$ and $c_{+}=0$, so we find that
\begin{equation}\label{52}
 E_{+-} =\frac{4c_{-}}{3l^{3}}r_{e}^{2}+\frac{1}{2 \mu l} \lim _{r \to \infty} \left(-r \dfrac{\partial H_{+}}{\partial r}+r^{2} \dfrac{\partial ^{2} H_{+}}{\partial r^{2}}\right),
\end{equation}
\begin{equation}\label{53}
 J_{+-}=\frac{4c_{-}}{3l^{2}}r_{e}^{2}+\frac{1}{2 \mu } \lim _{r \to \infty} \left( r\dfrac{\partial H_{+}}{\partial r} - r^{2} \dfrac{\partial ^{2} H_{+}}{\partial r^{2}}+\frac{l^{4}}{2 \mu r^{2}} \dfrac{\partial ^{2} H_{+}}{\partial t^{2}}\right) ,
\end{equation}
As the last case, suppose that $H_{-}(t,r)=0$ and $c_{-}=0$, then we have
\begin{equation}\label{54}
 E_{++} =\frac{4}{3l}\lim _{r \to \infty} \biggl\{ \left( H_{+}-\frac{1}{2}r\dfrac{\partial H_{+}}{\partial r} \right) c_{+}
+\frac{3r}{8 \mu} \left( - \dfrac{\partial H_{+}}{\partial r}+r \dfrac{\partial ^{2} H_{+}}{\partial r^{2}}\right) \biggl\},
\end{equation}
\begin{equation}\label{55}
 J_{++}=\frac{4}{3}\lim _{r \to \infty} \biggl\{ - \left( H_{+}-\frac{1}{2}r\dfrac{\partial H_{+}}{\partial r} \right) c_{+}
 +\frac{3r}{8 \mu} \left( \dfrac{\partial H_{+}}{\partial r}-r \dfrac{\partial ^{2} H_{+}}{\partial r^{2}}\right) +\frac{3l^{4}}{16 \mu r^{2}} \dfrac{\partial ^{2} H_{+}}{\partial t^{2}} \biggl\} ,
\end{equation}
In equations \eqref{50}-\eqref{55} we have just listed some special cases of equations \eqref{44} and \eqref{45}.

\section{Conclusion}
In this paper, we have studied the exact solutions of GMMG model in the non-linear regime of the theory and in its critical points. The critical points of the model in parameter space are the points where the central charge of dual CFT vanish. So we have a couple critical points correspond to the $c_-=0$, and $c_+=0$. The first type of exact non-linear solutions we have obtained are the pp-wave solutions in $AdS_3$ background. Then by going to the critical points correspond to the $c_-=0$, and $c_+=0$, we have obtained a logarithmic behavior for the profile of the wave. So in these critical points we have obtained new logarithmic modes. The presence of the logarithmic terms leads to divergences from the asymptotically
$AdS_3$ spacetime which, consequently, lead to declinations from the conformal symmetry on the boundary.
In order to restore the appropriate asymptotic behavior for the $AdS_3$ wave solutions, namely, its
Hamiltonian generators of the symmetries on the boundary to be represented by two
nontrivial copies of the infinite dimensional Virasoro algebra \cite{Brown:1986nw}, a relaxed set
of boundary conditions has to be imposed. Therefore, it is at the special points of the theory where
one has to impose the relaxed falloff conditions and also a LCFT arises
as dual to the background of the $AdS_3$ waves in the bulk \cite{Grumiller:2009sn,Grumiller:2009mw}.\\
Then we have studied the non-linear deformation of $AdS_3$  vacuum. We have considered time-independent solution and have shown that in the critical points not only we have logarithmic solution but also log-squared solution appeared. After that we have investigated logarithmic deformation of extremal BTZ (eBTZ) black hole solutions. For this case we have obtained both time independent and time-dependent solutions. Finally we have calculated the energy and angular momentum of logarithmic deformation of eBTZ black hole in critical points.

\section{Acknowledgments}
We would like to thank the anonymous referee for useful
comments on this work.

\end{document}